\documentclass[12pt,oneside,letterpaper]{article}
\usepackage{titlesec}
\titleformat*{\section}{\large\bfseries}
\titleformat*{\subsection}{\normalsize\bfseries}
\usepackage{amssymb}
\usepackage{amsmath}
\usepackage[dvips]{graphicx}
\usepackage{setspace}
\usepackage{amsfonts}
\usepackage{fancyhdr}
\usepackage{xcolor}
\usepackage{caption}
\usepackage{graphicx}
\usepackage{rotating}
\usepackage{comment}
\usepackage{color}
\usepackage{cite}
\usepackage{braket}
\definecolor{darkgreen}{rgb}{0,0.5,0}
\definecolor{darkblue}{rgb}{0,0,0.6}
\definecolor{purple}{rgb}{0.4,.2,0.7}
\newcommand{\p}{\partial}

\newcommand{\f}{\frac}

\newcommand{\be}{\begin{equation}}
\newcommand{\ee}{\end{equation}}

\usepackage[colorlinks=true,citecolor=darkgreen,linkcolor=black,urlcolor=purple]{hyperref}

\usepackage{pdfsync}

\makeatletter
\newcommand*{\defeq}{\mathrel{\rlap{%
                     \raisebox{0.3ex}{$\m@th\cdot$}}%
                     \raisebox{-0.3ex}{$\m@th\cdot$}}%
                     =} 
\makeatother

\def\be{\begin{eqnarray}}
\def\ee{\end{eqnarray}}

\newcommand{\bea}{\begin{eqnarray}}
\newcommand{\eea}{\end{eqnarray}}
\def\ben{\begin{equation}}
\def\een{\end{equation}}

\let\l=\lambda    \let\p=\phi \let\r=v

\let\f=\frac

\def\be{\begin{equation}}
\def\ee{\end{equation}}
\def\ba{\begin{array}}
\def\ea{\end{array}}

\def\ba#1\ea{\begin{align}#1\end{align}}
\def\bs#1\es{\begin{split}#1\end{split}}

\renewcommand{\p}{\partial}

\usepackage[top=1in, bottom = 1in, left = 1in, right = 1in]{geometry}

\allowdisplaybreaks  

\begin{document}
\onehalfspacing

\begin{center}

~
\vskip5mm

{\LARGE  {
Quantum gravity and the measurement problem\\ \vspace{2mm}in quantum mechanics
}}

\vskip8mm

Edgar Shaghoulian

\vskip8mm
{ \it UC Santa Cruz\\
Physics Department\\
1156 High Street\\
Santa Cruz, CA 95064}

\tt{eshaghoulian@ucsc.edu}

\end{center}

\vspace{4mm}

\begin{abstract}
\noindent
The measurement problem in quantum mechanics is almost exclusively discussed in situations where gravity is ignored. We discuss some recent developments in our understanding of quantum gravity and argue that they significantly alter the problem. Quantum gravity may even resolve one of the thorniest questions in discussions of the measurement problem: who collapses the wavefunction of the entire universe? 
\\
\\
\\
{\centering \emph{Essay written for the Gravity Research Foundation 2023 Awards for Essays on Gravitation.}}  
\end{abstract}



\thispagestyle{empty}
\pagebreak
\pagestyle{plain}

\setcounter{tocdepth}{2}
{}
\vfill
\clearpage
\setcounter{page}{1}

\section{Introduction}
The measurement problem in quantum mechanics has been around since the inception of quantum mechanics itself. It is concerned with understanding how our classical world emerges from fundamentally quantum-mechanical variables. While progress has been made, especially through the lens of decoherence,  foundational questions remain. The fact that these questions remain unsolved after almost 100 years of work is an indication that quantum mechanics is incomplete as a physical theory.  In this essay we propose that the incorporation of gravity will resolve these foundational issues and provide a complete physical theory. 

We will focus on one particularly startling feature of quantum gravity that influences the measurement problem in quantum mechanics. Through studying the consistency of quantum mechanics in black hole evaporation, we are naturally led to the proposal that the Hilbert space of a closed universe (i.e. one with no spatial boundary boundary) is one-dimensional. In such a situation, wavefunction collapse is not necessary: there is only one state the system could be in!\footnote{George Konstantinidis first emphasized to me how a one-dimensional Hilbert space would obviate the need for wavefunction collapse. At the time ($\sim$ 2014) we did not know of any interesting systems described by one-dimensional Hilbert spaces, but the entire universe is surely an interesting system.} This is especially powerful as a statement about the entire universe, because the crutch of an environment that collapses the wavefunction is unavailable. While we have not measured the spatial topology of our universe, at late times our universe is well-described by de Sitter space which can have closed spatial topology. 

In what follows we will review the measurement problem in quantum mechanics and the evidence that quantum gravitational effects imply a one-dimensional Hilbert space for a closed universe. We will end with a discussion of more general ideas in quantum gravity which impact the formulation of the measurement problem. 
\section{Foundations and interpretations of quantum mechanics}
Quantum mechanics provides deterministic evolution of the wavefunction through the Schr\"odinger equation
\be\label{schrod}
i \hbar \f{\p}{\p t } \Psi (x,t) = \left[-\f{\hbar^2}{2m} \f{\p^2}{\p x^2}+V(x,t)\right] \Psi(x,t)\,.
\ee
Physical observables correspond to Hermitian operators in a Hilbert space. The Born rule tells us how to predict outcomes of experiments that measure these physical observables: the possible outcomes are given by the eigenvalues $\lambda_i$ of the observable $\mathcal{O}$, and the probability of getting a particular $\lambda_i$ is given by $|\langle\Psi  |i\rangle|^2$, where $|i\rangle$ is the eigenvector corresponding to the eigenvalue $\l_i$.

This formulation of quantum mechanics often follows what may loosely be called the Copenhagen interpretation. The outcomes of observations are interpreted as being due to ``collapse" of the wavefunction: once an observable $\mathcal{O}$ is chosen to be measured, the measurement leads to the wavefunction of our system $|\Psi\rangle$ transitioning into one of the eigenvectors $|i\rangle$ of $\mathcal{O}$, with probability given by $|\langle \Psi|i\rangle|^2$. There are two unpalatable features of this interpretation. The first is that wavefunction collapse is not describable by Schr\"odinger evolution \eqref{schrod}, in particular it is not even time-reversible. This is often known as the problem of definite outcomes. The second is that the basis in which outcomes occur depends on what observable $\mathcal{O}$ we decide to measure. This is often known as the preferred basis problem. To deal with both of these problems, one would like to \emph{derive} the Born rule -- which encodes both preferred bases and definite outcomes -- from some more fundamental principles. 

Let's elaborate on the problem of a preferred basis. For this we need to review the notion of pre-measurement as described by von Neumann \cite{neumann}. This is a unitary process, governed by the Schr\"odinger equation, that describes two systems which become entangled with one another. We have our system $S$ in state $\sum_i c_i |s_i\rangle$ which comes into contact with our measurement apparatus $A$ initialized in state $|a_0\rangle$. Together they evolve unitarily as 
\be\label{super}
|\Psi\rangle = \left(\sum_i c_i |s_i\rangle\right)\otimes |a_0\rangle \longrightarrow \sum_i c_i |s_i\rangle \otimes |a_i \rangle\,.
\ee
Often the final form of the state is not unique, it can be rewritten in equivalent forms in different bases 
\be\label{super2}
|\Psi\rangle = \sum c_i' |s_i'\rangle |a_i'\rangle = \sum c_i'' |s_i''\rangle |a_i''\rangle = \cdots. 
\ee
For example, we can prepare an electron in a $\sigma_x$ eigenstate, which in terms of the basis of eigenvectors of $\sigma_z$ is given by 
\be
\frac{1}{\sqrt{2}}\left(|\uparrow_z\rangle + |\downarrow_z\rangle\right)
\ee
When it comes into contact with the measuring apparatus, initialized in state $|a_0\rangle$, they evolve into
\be
|\Psi\rangle = \f{1}{\sqrt{2}}\left(|\uparrow_z\rangle + |\downarrow_z\rangle\right) \otimes |a_0\rangle \longrightarrow \f{1}{\sqrt{2}} \left(|\uparrow_z \rangle \otimes |a_{\uparrow_z}\rangle + |\downarrow_z\rangle \otimes |a_{\downarrow_z}\rangle\right).
\ee
In the final state, the first term corresponds to the electron having spin up in the $z$ basis and the measuring apparatus registering spin up, and the second term corresponds to the electron having spin down in the $z$ basis and the measuring apparatus registering spin down. Let's say the measurement apparatus registering a particular spin is physically represented by another electron being in that same spin state. Then we would have
\be\label{way1}
\Psi =\f{1}{\sqrt{2}} \left(|\uparrow_z \rangle \otimes |\uparrow_z\rangle + |\downarrow_z\rangle \otimes |\downarrow_z\rangle\right).
\ee
But now we see that we can rewrite the same state in terms of eigenstates of the $\sigma_x$ operator
\be\label{way2}
\Psi =\f{1}{\sqrt{2}} \left(|\uparrow_x \rangle \otimes |\uparrow_x\rangle + |\downarrow_x\rangle \otimes |\downarrow_x\rangle\right).
\ee
Since we only ever observe particular outcomes on our measurement apparatus, only one of the terms in $|\Psi\rangle$ is selected. But is it one of the terms in \eqref{way1} or one of the terms in \eqref{way2}? It can't be both, since $\sigma_x$ and $\sigma_z$ are non-commuting operators. To make matters worse, in general the state can be written in an infinite number of ways, as in \eqref{super2}. Clearly, Schr\"odinger evolution by itself does not select the basis which determines outcomes of experiments. Before we can discuss why there is a unique term chosen at all, we need to understand what preferred basis the term is chosen in.

The program of decoherence \cite{Zeh:1970tq, zurek} has effectively solved the  preferred basis problem through the introduction of a monitoring environment. This leads to a preferred basis of classically stable states known as ``pointer" states. In short \cite{Zeh:1970tq, weinberg_2015}, we can \emph{choose} a basis of classical states, say one where two position eigenstates are separated by a kilometer. The role of the environment will be to introduce erratic phases into the wavefunction, leaving unperturbed the real part (since the weak coupling with the environment cannot shift the positions). Integrating over these random phases leads to a suppression of off-diagonal terms in the density matrix, with the diagonal terms being given by the classical basis states we began with. 

\begin{figure}
\centering
\includegraphics[scale=.5]{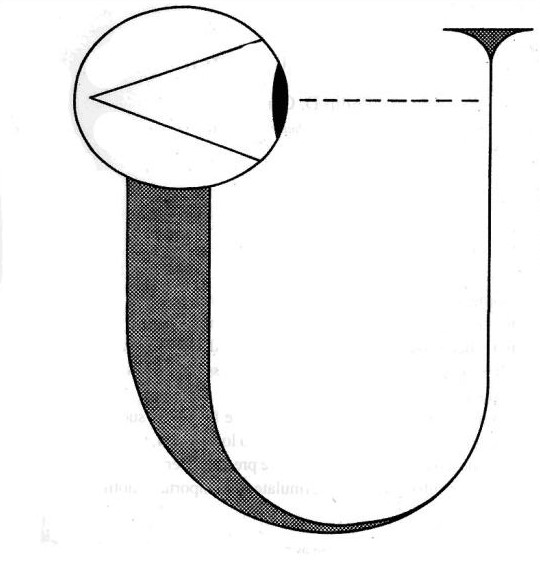}
\caption{Who collapses the wavefunction of the entire universe? \emph{Figure due to John Wheeler}  \cite{wheeler1980beyond}.}\label{wheeler}
\end{figure}

Even though decoherence solves the problem of the preferred basis, it requires the introduction of an environment. How does one understand quantum mechanics for our entire universe, where there is no monitoring environment? In other words, who collapses the wavefunction of the entire universe (Fig. \ref{wheeler})? Furthermore, the problem of definite outcomes -- why we experience just one of the terms in the superposition of the wavefunction -- has not been satisfactorily addressed by decoherence or any of the interpretations of quantum mechanics.\footnote{It has been argued, in particular by Zurek \cite{zurek}, that the measurement problem and the emergence of classicality is only a problem when resolving our universe into subsystems. This is because it is only as observers (subsystem 1) that we witness a world (subsystem 2) that behaves classically. While this is true, an argument for classicality on the scale of the entire universe may nevertheless shed light on classicality of subsystems within the universe. The consistent/decoherent histories approach \cite{Griffiths:1984rx, griffiths_2001, sep-qm-consistent-histories, Gell-Mann:1992wkv, omnes, Hohenberg:2010wmt} and the many worlds interpretation \cite{Everett:1957hd}  are concerned precisely with the wavefunction and classicality on the scale of the entire universe.}

There are many other interpretations that try to address these issues. One of the most popular is Everett's many worlds interpretation \cite{Everett:1957hd}, which says that the wavefunction never actually collapses. Instead, as systems interact with their environments, coherent superpositions are formed, and our lived reality corresponds to just one of the terms in the superposition. All the other terms are still there (so the wavefunction did not collapse) and correspond to the many other worlds in the many worlds interpretation. 

A large set of work concerns itself with modifying quantum mechanics with the addition of ``hidden variables" which alter ordinary Schr\"odinger evolution and lead to collapse of the wavefunction. Such attempts go beyond interpretations of quantum mechanics by modifying the dynamical equations themselves. 

There is \emph{much} more work on this subject than can reasonably be reviewed here. For a particularly illuminating reference, the reader is directed to \cite{Schloss2007}. 



\subsection{Instrumentalist vs realist approaches}
In discussions of the measurement problem it is useful to draw a distinction between instrumentalist and realist approaches (see \cite{weinberg_2015} for a discussion). In the instrumentalist approach, we only care about the outcomes of experiments, and therefore our theories need not describe what is going on with the people carrying out the experiments. In the realist approach, we care about everything: system, apparatus, environment, etc. The wavefunction of the entire universe is meaningful, beyond what it may say about the outcomes of experiments.


\section{Lessons from gravity}

In this section we will discuss what we have learned about quantum gravity in the past thirty years, including some recent developments in the past few years. These developments force us to reconsider the measurement problem in a world with gravity. 

The first set of developments, beginning in the early 1990's, is the idea of holography for quantum gravity \cite{tHooft:1993dmi, Susskind:1994vu}. This principle suggests that the dynamical degrees of freedom in quantum gravity are actually located on a boundary of space. This has had various concrete realizations within string theory \cite{Banks:1996vh, Maldacena:1997re, Itzhaki:1998dd}, most notably in the AdS/CFT correspondence. However, it immediately leads to the following confusion: in a universe with no spatial boundary -- such as de Sitter spacetime which gives a good model for the accelerated expansion in our universe -- where are the dynamical degrees of freedom to be located? A natural conclusion from the holographic perspective is that there are no dynamical degrees of freedom, i.e. that the Hilbert space of the entire universe has just one state \cite{susskind}. 

This idea has gained strong support in the past few years through the analysis of unitarity during black hole evaporation. By careful application of the gravity path integral, we have discovered a new rule for computing the entropy of a system $R$ in quantum gravity. This rule, sometimes known as the island rule, is given as \cite{Penington:2019npb, Almheiri:2019psf}
\be\label{island}
S_{QG}(R) = \text{min}_I\,\, \text{ext}_{\partial I}\left(S_{QFT}(R\cup I) + \frac{\text{Area}(\partial I)}{4G}\right).
\ee
Let's first define the objects appearing in the right-hand-side. $I$ is a codimension-two spacelike region that is spacelike separated from $R$. $S_{QFT}(R\cup I)$ is the von Neumann entropy of system $R\cup I $ as would be computed using quantum field theory in a fixed curved spacetime, whereas $S_{QG}(R)$ is the von Neumann entropy of system $R$ in the full quantum gravity theory. Area$(\partial I)$ is simply the area of the boundary of region $I$. 

The formula \eqref{island} says that when computing the entropy of some system $R$ in quantum gravity, we go looking for so-called ``island" regions $I$ which can be inputted into the formula on the right-hand-side. These island regions should lead to the quantity on the right-hand-side being an extremum with respect to variations of $\partial I$. Choosing $I = \emptyset$, i.e. not choosing to append some region $I$, is a valid choice, in which case $S_{QG}(R) = S_{QFT}(R)$. This means there is always at least one valid choice of $I$. In the event there are many choices of $I$, the minimization min$_I$ indicates that we minimize over the multiple choices. It is important that system $R$ be in a region where we can ignore the effects of gravity, but this point will not concern us. 

Let's choose system $R$ to be the Hawking radiation emitted from a black hole that was formed by a collapsing star in a pure quantum state. Early in the lifetime of the black hole, meaning before the Page time \cite{Page:1993df}, we find $S_{QG}(R) = S_{QFT}(R)$ since $I = \emptyset$ is the minimal choice. This leads to an entropy that increases monotonically as the black hole evaporates. At the Page time, which is precisely when an ordinary computation using quantum field theory in a fixed curved spacetime will conflict with unitarity \cite{Hawking:1976ra, Page:1993df}, a more minimal $I$ appears which covers the inside of the black hole. As a result, the entropy of the radiation sharply turns around and begins to decrease, ending at zero entropy when the black hole is done evaporating. This has provided a brand new perspective on the black hole information paradox. 

Two facts give us strong confidence in the validity of \eqref{island}. The first is that it can be derived from the perspective of the gravity path integral \cite{Penington:2019kki, Almheiri:2019qdq}. The second is that the output of the formula is completely in line with expectations of unitarity in black hole evaporation.

The formula \eqref{island} has an incredible implication, first discussed in \cite{Penington:2019npb, Almheiri:2019hni}. Let us say we have two systems, one of which is a closed universe (i.e. no boundary in space) that we want to entangle with some system $R$ living outside of the closed universe. Say we do this by preparing a semiclassical state where we have some qubits in our closed universe in Bell pairs with qubits in system $R$. When we compute the entropy of system $R$ using \eqref{island}, however, it turns out that we can choose an $I$ which is the entire closed universe. Since it has no boundary, it is trivially an extremum with respect to variations of $\partial I = \emptyset$. Furthermore, Area$(\partial I)/4G$ = Area$(\emptyset)/4G = 0$. And since $S_{QFT}(R\, \cup \,I) = 0$, there cannot be a more minimal choice of $I$ since we already have $S_{QG}(R) = 0$, which is the smallest a von Neumann entropy can be. This example illustrates that {\bf there is no way to entangle yourself with a closed universe} -- whenever you try, you will find that using the rule \eqref{island} leads to the region $I$ covering the entire universe and giving $S = 0$. The only logical conclusion is that the Hilbert space of the closed universe has only one state vector (since if it had two, say $|0\rangle_{\text{universe}}$ and $|1\rangle_{\text{universe}}$, then we could form a combination with two auxiliary states $|0\rangle_{\text{aux}}$ and $|1\rangle_{\text{aux}}$ of our system $R$ to obtain $|0\rangle_{\text{universe}}\otimes |0\rangle_{\text{aux}} + |1\rangle_{\text{universe}}\otimes|1\rangle_{\text{aux}}$, which is clearly an entangled state). So we have dim $\mathcal{H}_{\text{universe}} = 1$. This conclusion also seems necessary to resolve the factorization problem in AdS/CFT \cite{Marolf:2020xie} and is closely related to other constraints in quantum gravity \cite{McNamara:2020uza}.\footnote{It has been argued in the case of de Sitter space that even one state is too many \cite{Goheer:2002vf}. As discussed there and in \cite{Susskind:2021omt}, nonperturbative effects that break the de Sitter symmetries (which we expect to occur) invalidate those arguments.}

\section{The measurement problem in gravity}
In almost all discussions of the measurement problem, gravity is ignored.\footnote{A notable class of counterexamples includes \cite{Karolyhazy:1966zz, Diosi:1984wuz}, where gravity is proposed to be responsible for the collapse of the wavefunction.}  There are two common reasons for doing this. The first is that gravity is thought of as just another force to be quantized. In that case, quantum mechanics and the measurement problem appear for the quantum mechanics of this force. The second reason is that measurement questions can be phrased in situations where gravity is exceedingly weak, and therefore seemingly safe to ignore. What we have learned in the past thirty years is that the quantization of gravity does not follow the simple recipe of the other forces. Furthermore, na\"ive notions of effective field theory that would let us ignore the effects of gravity can fail spectacularly. 

One of the most dramatic repercussions is that the Hilbert space of a closed universe is trivial. This seems to be a remarkably lucky state of affairs, as the entire universe \emph{definitely} does not have an environment to collapse its wavefunction. If the Hilbert space only has one state, then the wavefunction is already collapsed! This means that the problems of preferred bases and particular outcomes are absent on the scale of the entire universe. While these problems are solved, this clearly introduces additional, perhaps more challenging problems. What is the meaning of a universe that only has one state, and how can it possibly account for all the complexity we see in our universe? 

A useful perspective is provided by topological field theories.  Consider Chern-Simons theory quantized on the spatial manifold $S^2$. The Hilbert space of states on this manifold is trivial, dim $\mathcal{H} = 1$. One way to see this is that the theory has no local operators, and the states on $S^2$ are in one-to-one correspondence with the local operators of the theory. Nevertheless, subsystems can have nonvanishing Hilbert spaces and entropies. While this seems like a contradiction, it is due to the fact that once a subsystem is chosen, some of the gauge invariance (in particular the gauge transformations that act nontrivially on the entangling surface) is relaxed. This has the effect of introducing additional degrees of freedom, known as ``edge modes," which provide a nontrivial Hilbert space for subsystems. Physically this can be thought of as cutting open the nonlocal operators of the theory (the Wilson lines) which traverse the entangling surface, and the additional degrees of freedom are charges that a Wilson line is allowed to end on. (See e.g.  \cite{Donnelly:2014gva}.) Gravity is expected to be an extreme version of this state of affairs, where the gauge-invariant nonlocal operators are not integrated over a curve as with Wilson lines, but instead integrated over all of spacetime. Relaxing the gauge invariance of the theory is also what we expect to do when discussing ``relational" observables, which necessarily refer to a particular observer. 

Another perspective is provided by the relevant algebras of observables in a closed universe. As discussed in \cite{Chandrasekaran:2022cip, Witten:2023qsv}, the algebra of observables without including an observer is trivial. This is due to the Hamiltonian constraint in a closed universe, which forces $H = 0$ in the physical space of states. This trivial algebra is analogous to the trivial Hilbert space of the closed universe. The inclusion of an observer, however, introduces a nontrivial algebra of observables, essentially because some of the gauge symmetry is broken by the presence of the observer. This is analogous to the edge modes discussed previously, or the rich physics that can exist for subsystems even if the full system is trivial. 

\section{Discussion}
We have discussed some dramatic effects which occur due to quantum gravity in a closed universe. These effects force us to rethink the measurement problem in quantum mechanics when gravity is involved. For the same reason that the net charge must vanish on a closed manifold due to the Gauss Law constraint, one needs $H = 0$ on a closed manifold due to the Hamiltonian constraint of general relativity. The introduction of an observer immediately impacts this in the same way that the introduction of an electron would impact the Gauss Law constraint. Thus, any discussion about outcomes of experiments as seen by observers cannot ignore the gravitational effect of the observer. A concrete instantiation of this logic is the appearance of nontrivial von Neumann algebras of observables for an observer in a closed universe \cite{Chandrasekaran:2022cip, Witten:2023qsv}, whereas not fixing to an observer would imply a trivial algebra. This fits nicely with the instrumentalist position. 

We may still choose a realist perspective and insist on asking questions about the wavefunction of the entire universe. This is often the position of the skeptic, determined to invalidate any interpretation of quantum mechanics or effect from decoherence. If our universe has a closed spatial topology,\footnote{In better-understood examples of quantum gravity like AdS/CFT it seems that topology cannot be measured by a linear operator in the quantum-mechanical description, see e.g. \cite{VanRaamsdonk:2010pw, Bao:2015nca, Berenstein:2016pcx, Berenstein:2017abm, Jafferis:2017tiu}.} then quantum gravity provides a beautiful resolution: there is only one state in the Hilbert space of the universe. There is no need for a monitoring environment to ``collapse" the wavefunction, since quantum gravity has already done it for us! 
This remarkable conspiracy suggests the following conjecture: \emph{the consistency of quantum mechanics on the scale of the entire universe requires that the spatial topology of our universe is closed.} It would be fascinating to develop the theory of subsystems in such a universe to ensure that we can match the richness and complexity we see as observers. The appearance of nonvanishing entropies and nontrivial algebras of observables gives a hint that this should be possible.

\section*{Acknowledgments}
I would like to thank George Konstantinidis, Raghu Mahajan, Philipp Strasberg, and Leonard Susskind for useful conversations. Special thanks to Konstantinidis for fun discussions about these questions and Mahajan for encouraging the conjecture in the final paragraph. 

\small
\bibliographystyle{ourbst}
\bibliography{gravitymeasurement}

\end{document}